# Cement Dust Exposure and Risk of Hyperglycemia and Overweight among Artisans and Residents Close to a Cement Factory in Sokoto, Nigeria

**[*1]Yahaya T, [1]Sani KA, [2]Oladele E, [1]Yawa E, Musa M, [3]Abubakar M, [4]Sulaiman R, [1]Bilyaminu M**

[1]Department of Biological Sciences, Federal University Birnin Kebbi, PMB 1157, Kebbi State, Nigeria
[2]Biology Unit, Distance Learning Institute, University of Lagos, Nigeria
[3]Department of Mathematics, Federal University Birnin Kebbi, Kebbi State, Nigeria
[4]Department of Biochemistry and Molecular Biology, Federal University Birnin Kebbi, Nigeria

**\*Correspondence:** yahayatajudeen@gmail.com; yahaya.tajudeen@fubk.edu.ng

## Abstract

The potential health risks of cement dust exposure are increasingly raising concern worldwide as the cement industry expands in response to rising cement demand. This necessitates the need to determine the nature of the risks in order to develop appropriate measures. This study determined the effects of cement dust exposure on the weight and blood glucose levels of people residing or working around a cement company in Sokoto, Nigeria. Demographic information was obtained using questionnaires from 72 participants, which included age, gender, educational level, exposure hours, occupation, and lifestyle. The blood glucose levels and body mass index (BMI) were measured using a Fine Test glucometer and a mechanical scale, respectively. The results showed that most of the people living around the cement company were middle-aged men (31-40; 42.06%) with a primary (33.33%) or secondary (45.83%) school education. It showed that 30 (41.69%) of the participants were overweight while 5 (6.94%) were obese. Additionally, 52.78% of the participants were diabetic while 31.94% were prediabetic. Participants that were exposed for long hours (> 15 hours per day) were the most diabetic (20% of the participants), followed by smokers (15%), and artisans (7%). It can be concluded that exposure to cement dust from the company increased the risk of overweight, obesity, and hyperglycemia among the participants. These health risks were worsened by daily long hours of exposure, smoking, and artisanal pollutant exposure. Human settlements and artisans should not be located near the cement company, and the company should minimize pollutant emissions.

## 1.0 Introduction

Cement is the most widely used construction material; it is primarily used to manufacture concrete [1]. Cement's widespread use can be attributed to the versatility of limestone, its basic raw material. It can also be attributed to its affordability, adhesive strength, and reliability compared to other materials used for making concrete [1]. Concrete made from cement is extensively used for building infrastructure and thus plays a vital role in gross domestic product (GDP), employment, and exports [2]. This has led to a surge in global cement production in the last few decades. As of 2022, the global cement market had reached 4.1 billion tons, up from about 1.40 billion tons in 1995 [3]. In Nigeria, local cement production capacity increased from 2 million tons in 2002 to 50 million tons in 2019 [4].The industry also employs around 2 million people, directly and indirectly, and has attracted $6 billion in investment, saving Nigeria around N210 billion ($1.3 billion) annually in foreign exchange equivalent [4].

Unfortunately, cement production processes consumed enormous energy, causing the release of high amounts of carbon dioxide ($CO_2$) into the environment [3]. The cement industry is responsible for about 5% of manmade $CO_2$ emissions, contributing around 4% to global warming [5]. Cement production also releases pollutants such as dioxin, sulfur dioxide, nitrogen dioxide, heavy metals, and particulate matter, among others [6]. Studies have linked cement dust exposure with diseases such as organ damage, hematological and biochemical abnormalities, skin problems, bone diseases, cardiovascular diseases, respiratory disorders, cancer,and even death [7]. Most of the health hazards of cement dust pollution are often caused by heavy metals due to their non-biodegradability, toxicity, and persistence. Heavy metals that are often detected in cement dust include chromium (Cr), lead (Pb), iron (Fe),copper (Cu), and cadmium (Cd) [8]. Considering the health consequences of cement dust pollution, there is a need for periodic assessment of workers and people living nearby todetect early signs of exposure, which may in turn result in reduced morbidity and mortality. This could create a sustainable growth where cement companies produce optimally without jettison the health of workers and residents.

In Sokoto, northwestern Nigeria, a cement company was commissioned to produce cement for the nation. In doing so, the industry employed many skilled and unskilled workers, boosting the economies of the state and the nation. Unfortunately, the company's operations pollute the environment, yet there is a paucity of documented information on the effects of the company's operations on people living nearby. Such information becomes necessary to enable policy makers to develop preventive and ameliorative measures which will in turn reduce morbidity and mortality. As a result, this study was conceptualized to determine the effects of cement dust emanating from the company on weight and blood glucose levels of people living nearby and the role played by demographic characteristics on the effects.

## 2.0 Materials and Methods

### 2.1 Description of the study site

The cement company is located in Kalambaina town, in the Wamakko Local Government Area of Sokoto State, Nigeria, at latitudes of 13.0059° N and longitudes 5.2476° E [9]. Kalambaina is about 6 km from Sokoto (the state capital) [10]. The inhabitants of Sokoto are mainly the indigenousHausa and Fulani, with some settlers comprising Yoruba, Igbo, Nupe, and other ethnic groups. Being situated in the dry Sahel, Sokoto State has a high daily temperature of about28 °C, and could exceed 45 °C during the dry season. The mean annual rainfall is about 629 mm, with the rainy season beginning in June and ending in October [10]. The state has several solid minerals, which include gold, limestone, gypsum, iron, and copper, among others.

### 2.2 Study Design

Descriptive-cross sectional study design was employed in this study to obtain information on cement dust exposure and risk of hyperglycemia and overweight among artisans and residents close to a cement factory in Sokoto, Nigeria.

### 2.3 Study Population

The study population comprised all people living or working nearby the cement company who consented were allowed to participate in the study. In addition, they must have been living in the study area for at least two years. Individuals who failed the inclusion criteria were excluded.

### 2.4 Sampling Techniques

Purposive sampling technique was adopted for the study for its suitability in dealing with the extensive nature of the required population and time constraints. Therefore, the criteria to be utilized will focus exclusively artisans and residents that are close (2Km away) to a cement factory in Sokoto state, Nigeria.

### 2.5 Study Instrument

Data were collected using semi-structured questionnaires. Every participant was either given a questionnaire or had the questionnaire read to them by the researcher. The glucose meter and mechanical scale instrument were used in obtaining data related to hyperglycemia and overweight.

### 2.6 Validity and reliability of the instrument

The sample of structured questionnaire was thoroughly scrutinized by research team before administered to the respondents. It was also subjected to pre-test reliability by administering the 10% of the sample size questionnaires and test the hyperglycemia and overweight instruments to a trial group at Bodinga, which is an are close to study area with similar communities and culture.

## 2.7 Data collection

A total of Seventy-two (72) questionnaires were distributed artisans and residents close to a cement factory in Sokoto, Nigeria. The participants were given questionnaires to fill out, which were structured into two sections A and B. Section A contains demographics of the participants which include sex, age, marital status, and education level. Section B contains the blood glucose levels and the body mass index (BMI).

## 2.8 Measurement of blood glucose

A glucose meter manufactured by the Fine Test with model number C001579 was used to measure the blood glucose levels. A test strip containing a drop of blood from each participant was placed in the glucose meter, and values in mg/dL were displayed on the screen of the instrument.

## 2.8 Calculation of body mass index (BMI)

A mechanical scale (Camry-BR9707) was used to measure the weight of the participants, while a wooden ruler was used to measure their heights. The BMI was then calculated usingthe equation below.

$BMI = Weight/Height^2 \ (Kg/m^2)$.

## 2.10 Ethics approval and consent to participate

This study was approved by the ethical committee of the Federal University Birnin Kebbi, Kebbi State, Nigeria. The guidelines for conducting research on humans as outlined by the committee were followed.Participants also gave informed consent before the study began.

## 2.11 Data analysis

Results were shown in percentages and frequency distribution tables.

## 3.0 Results

### 3.1 Gender distribution of the respondents

The gender distribution of the participants is revealed in Table 1. 50 (69.44%) of the participants were male, while 22 (30.56%) were female.

**Table 1: Gender distribution of respondents living around a cement factory in Sokoto, Nigeria**

| Gender | Frequency | Percentages (%) |
|--------|-----------|-----------------|
| Male | 50 | 69.44 |
| Female | 22 | 30.56 |
| Total | 72 | 100 |

### 3.2 Age distribution of the respondents

Table 2 shows the age groups of the respondents. 31 (42.06%) of the participants fall within the age group 31-40 years old, followed by the age group 20-30 years old with 21 (30.56%)representatives, the age group 51-60 years old with 3 (8.33%)members, and the age group >60 years old with 1(1.39%) member.

**Table 2: Age distribution of respondents living or working around a cement factory in Sokoto, Nigeria**

| Age (year) | Frequency | Percentage (%) |
|------------|-----------|----------------|
| 0-30 | 22 | 30.56 |
| 31-40 | 31 | 42.06 |
| 41-50 | 12 | 16.67 |
| 51-60 | 6 | 8.33 |
| >60 | 1 | 1.39 |
| Total | 72 | 100 |

### 3.3 Educational qualifications of the respondents

Figure 1 depicts the educational qualifications of the participants. Participants that had primary education made up 33.33% (24); those that had secondary education constituted 45.83% (33); and those with tertiary education consisted of 15 participants (20.83%).

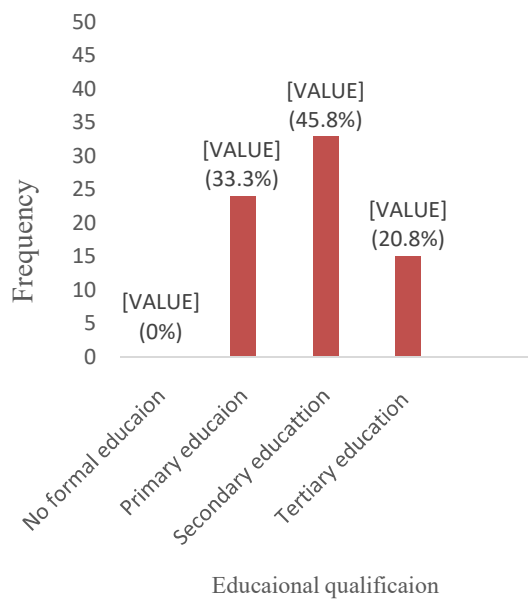

Figure 1: Educational qualifications of the respondents living or working around a cement factory in Sokoto, Nigeria.

### 3.4 Body mass index (BMI) of the respondents

Figure 2 displays the BMI of the participants. 37 (51.39%) of the study participants measured normal body weight, 30 (41.69%) were overweight, and 5 (6.94%) were obese.

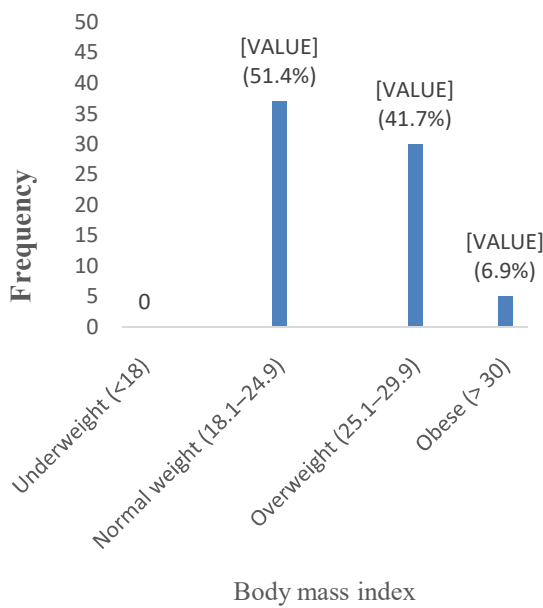

Figure 2: Body mass index (BMI) of the respondents living or working around a cement factory in Sokoto, Nigeria.

### 3.5 Blood glucose levels of the respondents

Figure 3 reveals the fasting blood glucose of the participants. 11 (15.28%) of the participants displayed normal blood glucose, 23 (31.94%) showed pre-diabetes, and 38 (52.78%) were diabetics.

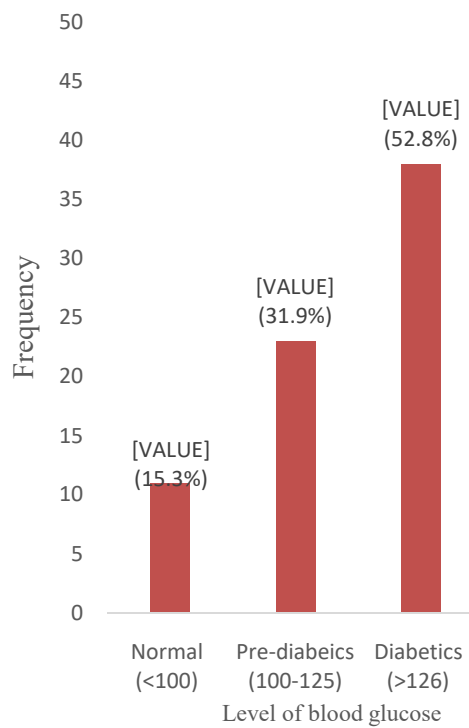

Figure 3: Fasting blood glucose levels of the respondents living or working around a cement factory in Sokoto, Nigeria

### 3.6 Effects of Life Styles on the blood glucose levels of the respondents

Figure 4 shows the effects of lifestyle as a co-factor on the blood glucose levels of the participants. 6 alcoholics, representing 8.33% of the respondents, were diabetic, while 20 non-alcoholics, representing 27.78%, were diabetic. 15 smokers (20.83%) were diabetic, while 8 non-smokers (11.11%) were diabetic.

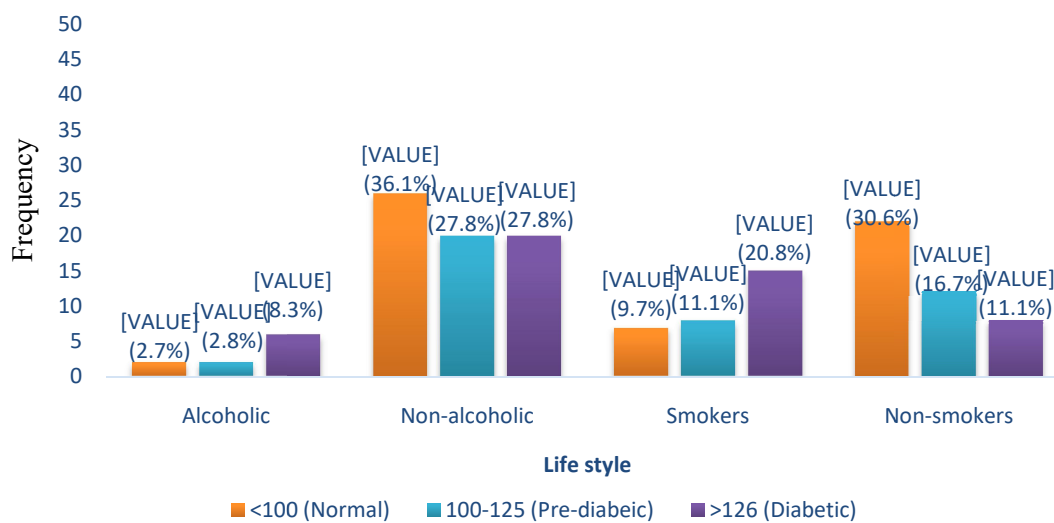

Figure 4: Blood glucose levels (based on life style) of the respondents living or working around a cement factory in Sokoto, Nigeria.

### 3.7 Effects of occupation on the blood glucose levels of the respondents

Table 3 reveals the blood glucose levels of the participants based on occupations. Artisans were the most diabetic, with 7 (9.72%) members, followed by farmers with 6 (8.33%) participants, traders with 3 (4.17%) participants, and civil servants with 2 (2.78%) participants.

**Table 3: Blood glucose levels (based on occupation) of the respondents living or working around a cement factory in Sokoto, Nigeria**

| Work | <100 | 100-125 | >126 | Percentage (%) |
|---|---|---|---|---|
| Farmers | 8(11.11%) | 7(9.72%) | 6(8.33%) | 29.17 |
| Traders | 7(9.72%) | 5(6.94%) | 3(4.17%) | 22.83 |
| Artisans | 9(12.50%) | 9(12.5%) | 7(9.72%) | 34.72 |
| Civil Servants | 6(8.33%) | 3(4.17%) | 2(2.78%) | 13.28 |
| Total | 72 | | | 100 |

Values were expressed as mg/dl; < 100 isnormal;100-125 ispre-diabetic;>125 isdiabetic

### 3.8 Influence of exposure time on the blood glucose levels of the respondents

The effect of exposure time on the blood glucose levels of the participants is revealed in Table 4. Participants who spent above 15 hours daily around the cement factory were the most diabetic, with 20 (27.78%) members, followed by those who spent between 8 and 15 hours with 6 (8.33%) members, and those who spent 0 to 7 hours with 5 (6.94%) member.

**Table 4: Influence of exposure time on the blood glucose levels of respondents living or working around a cement factory in Sokoto**

| Fasting blood glucose levels | | | | |
|---|---|---|---|---|
| Time (Hour) | <100 | 100 – 125 | >126 | Percentage (%) |
| 0 – 7 | 2(2.78%) | 4(5.56%) | 5(6.94%) | 15.28 |
| 8 – 15 | 5(6.94%) | 12(16.67%) | 6(8.33%) | 31.94 |
| Above 15 | 7(9.72%) | 11(15.28%) | 20(27.78%) | 52.78 |
| Total | 72 | | | 100 |

Values were expressed as mg/dl; < 100 =normal; 100-125 =pre-diabetic;        >125 =diabetic

## 4.0 Discussion

This study sought to determine the effects of a cement company's operations in Sokoto, Northwest, Nigeria, on the people living or working nearby using blood glucose levelsand body mass index (BMI) as health indices. Demographic data were collected and revealed that most of the respondents were males (69.44%), middle-aged (31-40; 42.06%), and mostly secondary school (45%) or primary school (33.0%)leavers. Most females stay indoors in the northern part of the country, which could be the reason more males than females consented to the survey. Middle-aged men and women are the most active and available members of any population, which could explain why their proportion was highest in this study. The preponderance of primary and secondary school leavers reflects the levels of education in that part of the country. Consistent with the findings of the current study, Oche et al. [11]reported the preponderance of male and middle-aged participants in a study carried out in Sokoto. Abubakar et al [12] also reported the preponderance of male participants in a study carried out in secondary schools in Sokoto. In a survey also carried out in Sokoto by Mohammed et al. [13], male and secondary school participants were the most dominant. Moreover, Awosanet al. [14] reported the dominance of middle-aged participants in a study carried out among tertiary health workers in Sokoto.

The study further showed that the majority of the participants (41.67%) were overweight, and (6.94%) were obese. These incident rates are much higher than the 27.2% prevalence of overweight and lower than the 10.4% prevalence of obesity in the northwest region (where the current study was conducted) reported in a systematic review by Chukwuonye et al. [15]. They are also much higher than the 26.0% prevalence of overweight and lower than the 15.0% obesity prevalence reported for Nigeria in a systematic review by Ramalan et al. [16]. Some pollutants that are often present in cement dust, such as heavy metals, can cause overweight and obesity by disrupting the dopamine signaling in the hypothalamus, causing oxidativestress in endoplasmic reticulum, thus impairing adipogenesis and adipocytokine synthesis [17,18]. An increase in body weight following the exposure of some people to cement dust was reported by Manjula et al. [19], which aligns with the current study. Merenuet al. [20], and Yahaya et al. [21], also observed weight increases among cement workers and rats exposed to cement dust. According to Yahaya et al [21], calcium in cement could be responsible for the increase in the weight of rats exposed to cement dust because it is a basic animal nutrient that is involved in bone and blood formation. However, in a study carried out by Owonikoko et al. [22], cement dust exposure decreased the weights of exposed organs in rats before histopathological damage. Meo et al. [23], also reported a lower BMI among cement factory workers in Saudi Arabia compared with the control group.

The prevalence of diabetes mellitus (52.78%) and prediabetes (31.94%) recorded among the study participants were higher than the 4.3% diabetes prevalence estimated for Nigeria by the WHO [24] and the 13.2% prediabetes recorded for Nigeria in a systematic review by Bashir et al. [25]. The results of the current study align with Meo et al. [23], who reported a 42.47% prevalence of diabetes and a 15.05% prevalence of prediabetes among cement factory workers in Saudi Arabia. Jung et al. [26], also reported an increased risk of diabetes mellitus among people living close to a cement plant in Jangseong-gun, South Korea. Cement contains heavy metals, which may accumulate in the liver and pancreas and disrupt gluconeogenesis and destroy insulin-producing beta cells, culminating in hyperglycemia and diabetes mellitus [27,28]. Heavy metals can also predispose humans to diabetes by causing insulin resistance, oxidative stress, and genetic or epigenetic changes in the beta cells [29]. Pollutants can also increase BMI (also noticed in the current study) and thus increase the risk of being diagnosed with diabetes mellitus [30]. In addition, cement production releases air pollutants such asnitrogen dioxide, carbon monoxide, particulate matter, and sulfur dioxide,all of which have been linked with insulin resistance and beta cell dysfunction via oxidative stress [31].

Diabetes was more prevalent among those who spent more time (> 15 hours) daily around the cement factory, as well as among smokers and artisans. These findings suggest that the duration of exposure, lifestyle, and occupation boost the health risks of cement dust exposure. In a study by Meo et al. [32], those who spent the most time around a cement factory in Saudi Arabia developed the worst health hazards. Duration-dependent effects were also reported among cement factory workers in Jordan [33]. Pollutants accumulate in the body with increasing durations of exposure, resulting in more health damage. Regarding smoking, substances in cigarettes such as nicotine and heavy metals can cause inflammation and oxidative stress, which are risk factors for diabetes mellitus [34]. Some artisanal works like welding and panel beating release poisonous substances, which may influence the development of hyperglycemia and diabetes. In a study of hyperglycemia prevalence in Lagos, Nigeria, by Yahaya et al. [35], artisanal workers and smokers were the most hyperglycemic.

## 5.0 Conclusion

The results showed that most of the people living or working around the cement company are middle-aged men (31-40) with a primary or secondary school education. More than half of the participants were either overweight (41.67%) or obese (6.94%). More than half of the participants (52.78%) were diabetic, and (31.94%) were prediabetic.Tobacco smoking, daily long hours of exposure, as well as artisanal pollutant exposure contribute to the high prevalence of diabetes mellitus and prediabetes in the study population. Over all, the results showed that operations at the cement company have negative effects on the residents. An environmental impact assessment (EIA) of any proposed cement factory must be conducted and pollution management strategies provided before the factory becomes operational. Public health and environmental officers should keep close watch on the cement factory to comply with environmental pollution management guidelines. People are advised not to live near a cement factory.More studies like the current one is advised.

## 6.0 Conflict of Interest

Authors declared no conflict of interest.